# Elemental characterization of the Avogadro silicon crystal WASO 04 by neutron activation analysis

G. D'Agostino(1), L. Bergamaschi(1), L. Giordani(1), G. Mana(2), E. Massa(2) and M. Oddone(3)

(1) Istituto Nazionale di Ricerca Metrologica (INRIM), Unit of Radiochemistry and Spectroscopy, c/o Department of Chemistry, University of Pavia, via Taramelli 12, 27100 Pavia, Italy

(2) Istituto Nazionale di Ricerca Metrologica (INRIM), Strada delle Cacce 91, 10135 Torino, Italy.

(3) Department of Chemistry, University of Pavia, via Taramelli 12, 27100 Pavia, Italy

Email: g.dagostino@inrim.it

**Abstract:** Analytical measurements of the $^{28}$Si crystal used for the determination of the Avogadro constant are essential to prevent biased results or under-estimated uncertainties. A review of the existing data confirms the high-purity of silicon with respect to a large number of elements. In order to obtain a direct evidence of purity, we developed a relative analytical method based on neutron activation. As a preliminary test, this method was applied to a sample of the Avogadro crystal WASO 04. The investigation concerned twenty-nine elements. The mass fraction of Au was quantified to be $(1.03 \pm 0.18) \times 10^{-12}$. For the remaining twenty-eight elements, the mass fractions are below the detection limits, which range between $1 \times 10^{-12}$ and $1 \times 10^{-5}$.

## 1. Introduction

The determination of the Avogadro constant, $N_A$, through the x-ray crystal density molar mass method is a route for the redefinition of the unit of mass in terms of a fundamental constant [1]. To achieve a relative standard uncertainty of $3 \times 10^{-8}$, it was necessary to use a perfect silicon single crystal highly enriched with the $^{28}$Si isotope [2, 3]. To achieve this uncertainty, the total impurities within the $^{28}$Si crystal must not exceed a mass fraction of a few parts in $10^9$. In case of a higher value, the relevant mass fractions must be quantified and the $N_A$ value corrected for. The correction uncertainty must not exceed the mass fraction limit given above. There is a general consensus that, in silicon materials, the contaminations by the most of the natural elements are significantly smaller than parts per billion of a silicon atom. However, the collection of experimental data concerning the largest number of impurities, and directly related to Avogadro crystals, is worth to decrease the risk of biased results or of under-evaluated uncertainties.

Research in semiconductor technology has focused on analyses of silicon materials based on Neutron Activation Analysis (NAA). Starting from a review of the existing data and taking advantage of the multi-elemental feature and high sensitivity of neutron activation analysis, we developed an analytical method for the determination of trace elements in $^{28}$Si crystals. As a preliminary test, this method was applied to investigate the purity of the silicon single crystal WASO 04. The results are given in the present paper.

This natural silicon crystal was grown with nitrogen doping and purified by application of the float-zone technique by Wacker Siltronic in 1995. It was used in the last natural-silicon determination of $N_A$ [4] and as a precursor to develop and to demonstrate the technologies subsequently refined and used to determine the Avogadro constant by using a crystal highly enriched with $^{28}Si$. A partial elemental characterization of the this crystal can be found in [5]. In particular, trace impurity measurements, carried out by infrared absorption spectrometry, quantified atomic fractions of $5.8 \times 10^{-8}$, $2.6 \times 10^{-8}$ and $1.4 \times 10^{-8}$ for C, O and N, respectively, and reached atomic fraction detection limits of about $6 \times 10^{-9}$ for B and P.

The present paper is made up of 4 sections. Section 2 reports NAA results for silicon available in the literature. Section 3 recalls the measurement model of neutron activation. Section 4 describes the analysis of the WASO 04 crystal, and, finally, section 5 shows the experimental results.

## 2. Literature data

The first step for the production of monocrystalline silicon is the decomposition of high quality sand ($SiO_2$) into metallurgical-grade silicon and carbon monoxide by a carbothermic reaction. The second step is a purification which usually includes a distillation resulting in a silicon compound having impurity levels of a few ppb. The purified silicon compound is then used to deposit polysilicon onto a thin monosilicon seed that serves as a starting material in a chemical vapour deposition reactor. In the fourth and final step, the polysilicon is used to grow monocrystalline silicon ingots by the Czochralsky (CZ) or float-zone (FZ) method.

In the CZ method, the polysilicon material is melted in a crucible and a single crystal seed is used to start the crystal growth. Differently, in the FZ method, a radio frequency coil is used to zone-melt the polysilicon material. Thus, during the growth of a FZ crystal, the melt zone never comes into contact with anything but vacuum (or inert gases). In addition, multiple passages of the coil along the silicon ingot help to further remove the impurities by segregation and evaporation of most of the contaminants. The impurities in silicon materials can therefore originate from the residual elemental content of the original $SiO_2$ or can be added during the production process.

Neutron activation analysis has been widely used as a diagnostic technique in semiconductor industry. Literature data include several works reporting on residual impurities in semiconductor silicon determined by neutron activation. The review of the existing data provides an overview of the expected impurities. Tables 1, 2 and 3 report the atomic fractions of 20, 13 and 11 elements measured in bulks of polycrystalline silicon, FZ, and CZ grown single-crystal silicon, respectively. Data are given in chronological order with the most recent being reported in the last right hand column. The references from which data are taken are also given.

With the exception of Cu in FZ processed crystals, the average content of impurities decreased during the years both in polycrystalline and in single-crystal (FZ and CZ) silicon materials. A small fraction of the impurities from the crucible can be incorporated into the crystal [6] and, therefore, FZ crystals should be less contaminated than CZ crystals [7]. However, Bottger showed that the purity level of FZ and CZ crystals became similar during the years [8]. It is also worth noting that the literature data do not show a dependence of the overall impurity content on the crystal type (n-type or p-type) or, if doped, on the doping element [9]. Overall, the reported data highlight that, independently on the processing year and on the silicon material, only atomic fractions of Na, Fe and Cu reached the ppb level, whereas atomic fractions of Ag, As, Au, Br, Ce, Co, Cr, Eu, La, Mo, Sb, Ta, Tb, Th, U, W, Zn were always below the ppb level.

| | 1977 [10] | 1981 [11] | 1983 [8] | 1985 [8] | 1991 [9] | 1992 [12] |
|---|---|---|---|---|---|---|
| Ag | | | | | $1.3 \times 10^{-10}$ | |
| As | | $4.3 \times 10^{-11}$ | | | $8.6 \times 10^{-12}$ | $1 \times 10^{-11}$ |
| Au | | $3 \times 10^{-12}$ | | | $2.2 \times 10^{-13}$ | $7 \times 10^{-14}$ |
| Br | | | | | $1.5 \times 10^{-11}$ | |
| Ce | | | | | $4.5 \times 10^{-12}$ | |
| Co | $6 \times 10^{-11}$ | $4.5 \times 10^{-11}$ | $8 \times 10^{-13}$ | $< 1.4 \times 10^{-12}$ | $2.1 \times 10^{-11}$ | |
| Cr | $1.4 \times 10^{-10}$ | $< 0.1 \times 10^{-9}$ | $< 8 \times 10^{-12}$ | $1.2 \times 10^{-10}$ | $1.2 \times 10^{-11}$ | $6 \times 10^{-12}$ |
| Cu | $2 \times 10^{-11}$ | $1.8 \times 10^{-9}$ | $4 \times 10^{-12}$ | $< 6 \times 10^{-12}$ | | |
| Eu | | | | | $1.2 \times 10^{-11}$ | |
| Fe | $6 \times 10^{-9}$ | $7 \times 10^{-9}$ | $8 \times 10^{-10}$ | $1 \times 10^{-9}$ | $1.5 \times 10^{-9}$ | |
| La | | | | | $5.8 \times 10^{-12}$ | |
| Mo | $4 \times 10^{-11}$ | | | | | |
| Na | | $3.2 \times 10^{-9}$ | | | $9.5 \times 10^{-11}$ | $3 \times 10^{-10}$ |
| Sb | | $< 6 \times 10^{-12}$ | | | $1.4 \times 10^{-11}$ | $7 \times 10^{-13}$ |
| Ta | | | | | $9.2 \times 10^{-12}$ | |
| Tb | | | | | $6.8 \times 10^{-12}$ | |
| Th | | | | | $3.9 \times 10^{-12}$ | |
| U | | | | | $2.2 \times 10^{-11}$ | |
| W | $4 \times 10^{-11}$ | | | | $1.6 \times 10^{-11}$ | |
| Zn | | $8 \times 10^{-10}$ | | | $1.8 \times 10^{-11}$ | |

**Table 1.** Atomic fractions of impurities in polycrystalline Si.

| | 1977 [10] | 1979 [7] | 1983 [8] | 1985 [8] | 1991 [13] |
|---|---|---|---|---|---|
| Ag | | | | | $2 \times 10^{-11}$ |
| As | | | | | $1.3 \times 10^{-11}$ |
| Au | | $1.1 \times 10^{-11}$ | | | $5 \times 10^{-13}$ |
| Br | | | | | $1 \times 10^{-11}$ |
| Co | $2 \times 10^{-11}$ | $8 \times 10^{-12}$ | $2 \times 10^{-11}$ | $6 \times 10^{-13}$ | $< 9 \times 10^{-13}$ |
| Cr | $4 \times 10^{-11}$ | | $1.4 \times 10^{-11}$ | $8 \times 10^{-12}$ | $< 1 \times 10^{-11}$ |
| Cu | $2 \times 10^{-11}$ | $5 \times 10^{-10}$ | $1.2 \times 10^{-11}$ | $9 \times 10^{-11}$ | |
| Fe | $14 \times 10^{-9}$ | | $5.4 \times 10^{-10}$ | $< 1.4 \times 10^{-10}$ | $< 6 \times 10^{-10}$ |
| Mo | $1.4 \times 10^{-11}$ | | | | |
| Na | | $6 \times 10^{-10}$ | | | |
| Sb | | $< 6 \times 10^{-12}$ | | | $2 \times 10^{-12}$ |
| W | $1.2 \times 10^{-11}$ | | | | $< 3 \times 10^{-12}$ |
| Zn | | | | | $< 1.5 \times 10^{-11}$ |

**Table 2.** Atomic fractions of impurities in FZ Si.

| | 1977 [10] | 1979 [7] | 1981 [11] | 1983 [8] | 1985 [8] |
|---|---|---|---|---|---|
| As | | | $< 8 \times 10^{-12}$ | | |
| Au | | $2.3 \times 10^{-10}$ | $< 8 \times 10^{-13}$ | | |
| Co | $2 \times 10^{-11}$ | $1 \times 10^{-11}$ | $1 \times 10^{-11}$ | $2 \times 10^{-11}$ | $6 \times 10^{-13}$ |
| Cr | $1.2 \times 10^{-11}$ | | $< 1 \times 10^{-10}$ | $1.4 \times 10^{-11}$ | $8 \times 10^{-12}$ |
| Cu | $2 \times 10^{-10}$ | $2 \times 10^{-9}$ | $2 \times 10^{-10}$ | $5.6 \times 10^{-11}$ | $1.7 \times 10^{-11}$ |
| Fe | $14 \times 10^{-9}$ | | $< 1.6 \times 10^{-9}$ | $3 \times 10^{-10}$ | $4.4 \times 10^{-10}$ |
| Mo | $1.6 \times 10^{-11}$ | | | | |
| Na | | $1.3 \times 10^{-9}$ | $2 \times 10^{-10}$ | | |
| Sb | | $7 \times 10^{-11}$ | $< 6 \times 10^{-12}$ | | |
| W | $2 \times 10^{-12}$ | | | | |
| Zn | | | $< 3 \times 10^{-10}$ | | |

**Table 3.** Atomic fractions of impurities in CZ Si.

## 3. Measurement model of neutron activation analysis

The neutron activation analysis is based on the detection of the gamma rays emitted by radioactive isotopes produced by neutron bombardment of stable nuclei. When a non-elastic collision takes place between a neutron and the nucleus of a target isotope, a compound nucleus is formed in an excited state. This step is followed by an extremely rapid de-excitation to a more stable configuration, usually accompanied by the emission of prompt gamma rays. The new nucleus is usually radioactive and will de-excite by emitting delayed gamma rays or particles. In the latter case, the resulting nucleus is often still excited and a further gamma emission could occur. The energy spectrum of the emitted gamma rays shows discrete energy peaks which identify the radioactive nucleus.

In more details, neutron activation analysis can be described as follows. When a neutron beam traverses a target consisting of $N$ identical target nuclei, a small fraction of neutrons will react with them. The ratio between the number of reactions per target nucleus, $n/N$, per unit time and neutron flux, $\Phi$, i.e., the number of incident neutrons per unit area per unit time, is the cross section $\sigma$ for the reaction. The measurement unit commonly adopted for $\sigma$ is the barn (1 b = $10^{-24}$ $\mathrm{cm}^2$). The cross section depends on the neutron energy and therefore the energy spectrum of incoming neutrons affects the number of reactions.

Since the de-excitation takes place also during the irradiation, according to the law of disintegration of radioactive nuclei having half-life $t_{1/2}$, the number of radioactive nuclei expected in a sample after a neutron bombarding lasting a time $t_\mathrm{i}$ is

$$n(t_\mathrm{i}) = \frac{\Phi\, \sigma\, N}{\lambda}(1 - e^{\lambda t_\mathrm{i}}), \tag{1}$$

where $\lambda = \ln(2)/t_{1/2}$ is the decay constant. Equation (1) neglects the burn-out of the target nuclei during the irradiation, i.e., the reduction of target nuclei due to the production of compound nuclei. The error due to this approximation is commonly negligible because the reduction rate per target nucleus $\Phi\,\sigma$ is, in practice, always lover than $10^{-9}$ $\mathrm{s}^{-1}$ when the bombardment is performed with low-power research neutron sources. Equation (1) neglects also the self-shielding. This effect can be significant and, to some extent, depends on sample size and cross sections of the matrix elements.

As soon as the sample is removed from the neutron flux, the production of compound nuclei stops and the number of radioactive nuclei decays according to the law of disintegration. After a time $t_\mathrm{d}$ from the end of the irradiation, the number of radioactive nuclei inside the sample is

$$n(t_\mathrm{d}) = n(t_\mathrm{i}) e^{-\lambda t_\mathrm{d}}. \tag{2}$$

To identify and to count the number of activated nuclei, the number of gamma photons having the same energy $E$,

$$n_\gamma = n(t_\mathrm{d})(1 - e^{-\lambda t_\mathrm{c}})\Gamma_E \tag{3}$$

is counted during the time interval $t_\mathrm{c}$. In (3), $\Gamma_E$ is the radiation yield. Equation (3) neglects the self-absorption of gamma rays occurring within the sample. As for the self-shielding, the self-absorption mainly depends on sample size and cross sections of the matrix elements.

Actually, only the fraction $\varepsilon$ of the emitted gamma photons is collected and the corresponding counts, $n_c$, are stored by the multichannel analyzer according to their energy. The fraction $\varepsilon$ is the detection full-energy peak efficiency; it depends on several parameters, such as the energy of the gamma rays, the distance between the sample and detector, the size and shape of the sample and detector.

Since the total count $n_t$ stored by the multichannel analyzer includes also a background count $n_b$, the net count $n_c$ is obtained by difference. Moreover, during the counting time, there are dead times of the detection system due to the processing times of the electronics. It is convenient to define a live time, $t_{live}$, of the detection system, i.e., the counting time less the sum of the dead times. Under the hypothesis of a constant count rate during the data collection, the stored counts are corrected by a factor $t_c / t_{live}$. It follows that

$$(n_t - n_b)\frac{t_c}{t_{live}} = n_\gamma\, \varepsilon \,. \tag{4}$$

Therefore, from eqs. (1), (2), (3) and (4), the amount of target nuclei $N$ of an isotope in a sample that emits gamma rays after activation in a neutron flux $\Phi$ can be estimated according to

$$N = \frac{(n_t - n_b)}{(1 - e^{\lambda t_i})\, e^{\lambda t_d}\, (1 - e^{\lambda t_c})}\, \frac{\lambda}{\Phi\, \sigma\, \Gamma_E\, \varepsilon}\, \frac{t_c}{t_{live}} \,. \tag{5}$$

If the isotopic fraction, $\theta$, of the target isotope is known, the number of atoms of the corresponding element is also known.

Although (5) shows that the knowledge of $\Phi$, $\sigma$, $\Gamma_E$ and $\varepsilon$ allows an absolute determination of $N$ to be made, neutron activation analysis is commonly used in a relative measurement protocol using standards of pure elements. In the relative method, both the sample and the standards are co-irradiated in a similar neutron flux. Since the number of target nuclei in the standard, $N_{st}$, is known, the number of target nuclei within the sample is computed according to

$$N = \frac{\dfrac{(n_t - n_b)}{e^{\lambda t_d}\, (1 - e^{\lambda t_c})}\, \dfrac{1}{\varepsilon}\, \dfrac{t_c}{t_{live}}}{\left.\dfrac{(n_t - n_b)}{e^{\lambda t_d}\, (1 - e^{\lambda t_c})}\, \dfrac{1}{\varepsilon}\, \dfrac{t_c}{t_{live}}\right|_{st}}\, N_{st} \,, \tag{6}$$

where the st subscript indicates that all the quantities in the denominator refer to the standard. Model (5) highlights the independence of the results of the analysis on the physical state of the sample. Moreover, the simultaneous collection of counting peaks corresponding to different radioactive nuclei makes the neutron activation a multi-element technique.

## 4. Determination of the impurities in the Avogadro Si crystal WASO 04

Since silicon isotopes either produce short-lived radionuclides or have low activation cross sections, silicon is suitable for Instrumental Neutron Activation Analysis (INAA), i.e., activation analysis without chemical destruction of the sample.

In natural silicon, the main neutron capture reactions are $^{30}$Si(n, $\gamma$)$^{31}$Si and $^{29}$Si(n, p)$^{29}$Al, that is, the stable isotopes $^{30}$Si and $^{29}$Si are transformed in the radioactive isotopes $^{31}$Si and $^{29}$Al. The relevant cross sections are not negligible, but the half-lives of $^{31}$Si and $^{29}$Al are about 2.6 h and 6.6 min. Thus, after two or three days, the bulk activity is drastically reduced and the gamma emission of the impurities producing isotopes with a longer half-life can be measured. The only medium-lived radionuclide produced by pure silicon is $^{24}$Na, which has 15 h half-life and it is produced by $^{28}$Si(n, $\alpha$p)$^{24}$Na [10]. This reaction interferes with the determination of Na, which is carried out by measuring the gamma emission of $^{24}$Na produced by neutron activation of the isotope $^{23}$Na through the reaction $^{23}$Na (n, $\gamma$) $^{24}$Na. Therefore, though the $^{28}$Si(n, $\alpha$p)$^{24}$Na reaction requires fast neutrons and the relevant cross section is extremely small (2 nb), the contamination due to Na could be overestimated.

Figure 1 shows a picture of the sample of the WASO 04 crystal used for this study. The sample was taken from the WASO 04 crystal at the axial position 90 cm, according to the reference frame described in [5]. Diameter, length and weight are 10 mm, 34 mm and 6 g, respectively.

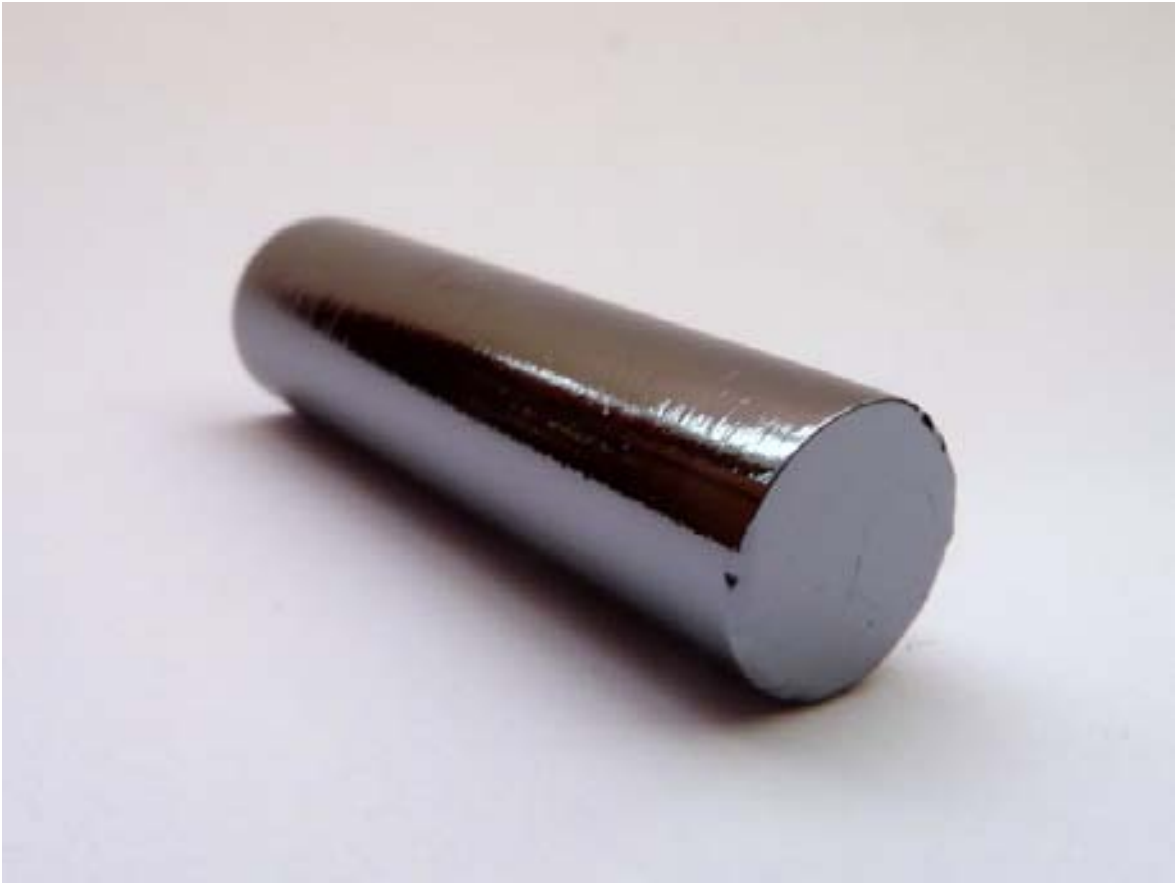

**Figure 1.** The WASO 04 sample.

The determination of impurities was carried out according to the model (6). Reference solutions and materials were used as standards. In particular, two standards, ML and LL, were gravimetrically prepared by pipetting aliquots of multi-element solutions onto filter papers rolled up as a cylinder and inserted in polyethylene vials (Kartell®, 1 mL). Before sealing the vials, the pipetted aliquots were evaporated to dryness using an infrared lamp in a fume hood under ambient conditions. The multi-element solutions were home-made by mixing certified single-element solutions to suitable mass fractions. In addition, two more standards, SJ and BR, were prepared by weighting and sealing aliquots of reference materials in polyethylene vials (Kartell®, 1 mL).

For irradiation, the WASO 04 sample was sealed in a polyethylene vial (Kartel®, 8 mL) and packed in a container together with the four standards. The neutron irradiation lasted 6 hours and was performed in the central thimble of the 250 kW TRIGA Mark II reactor at the Laboratory of Applied Nuclear Energy (LENA) of the University of Pavia. The thermal and epithermal neutron fluxes were about $6 \times 10^{12}$ cm$^{-2}$ s$^{-1}$ and $5.5 \times 10^{11}$ cm$^{-2}$ s$^{-1}$, respectively. At the end of the irradiation, the sample and standards were left to cool until the activity of $^{31}Si$ decayed to safe values.

Before counting, the standards were rinsed with pure water and diluted nitric acid in order to remove any adhering contaminations of the vials. The silicon sample was taken away from its irradiation vial and a deep surface layer was removed. This eliminated either the impurities in the silicon surface layer and the possible deposition of contaminants on the surface occurred during preparation and handling. Accordingly, the silicon sample was washed with trichloroethylene, acetone and deionized water, etched for 25 min with a solution 10:1 of nitric acid (assay 67-69%) and hydrofluoric acid (assay 47-51%), and finally rinsed in deionized water, ethylalcohol and acetone. The loss of mass during etching was about 270 mg. Lastly, the silicon sample was sealed in a new polyethylene vial for counting.

Gamma spectrometry was performed with an automatic system comprising a sample changer, a coaxial HPGe detector ORTEC® GEM50P4-83 (relative efficiency 50%, resolution 1.90 keV FWHM at 1332 keV), a digital signal processor ORTEC® DSPEC jr 2.0, and a personal computer running a software for data acquisition and processing ORTEC® Gamma Vision 6.0. Calibrations of energy and of detector resolution and efficiency were carried out by a standard multi-gamma source LEA 9ML01EGMA. The calibrations were performed in two counting positions, 0 and 8, by placing the standard source in contact with and 8 cm far from the head of the detector, respectively.

The gamma spectra were recorded in three different periods. The first was recorded after 3 days cooling of both the WASO 04 sample and ML standard; the counting times were 2 h and 8 h at the counting positions 0 and 8, respectively. The second spectrum was recorded after 7 days cooling of the SJ and BR standards; the counting times were 6 h at the counting position 8. Finally, the last spectrum was recorded after 16 days cooling of the LL standard; the counting time was 24 h at the position 8.

Since the standards were gravimetrically prepared, the number of target nuclei of an isotope in the standard is

$$N_{\mathrm{st}} = \frac{\theta\, m}{M} N_{\mathrm{A}}, \tag{8}$$

where $\theta$ is the isotopic fraction, $m$ and $M$ are the mass and the atomic mass of the standard element, and $N_{\mathrm{A}}$ is the Avogadro constant. The number of target nuclei of the standard element in the sample is therefore $N/\theta$, where $N$ is calculated according to model (6).

## 5. Results

The results expressed in terms of both the mass fractions and atomic fractions of twenty-nine elements in the WASO 04 sample are shown in table 4. The values for Nd and Yb are only indicative because the corresponding mass fractions in the SJ standard are not certified. The standard, the half-life of the radioactive nucleus, the neutron capture reaction and the energy peak are also reported. Standard uncertainties include only the component due to the counting statistics and the detection limits are evaluated according to the Currie's method [14].

It was possible to quantify only the mass fractions of $^{198}Au$ and $^{24}Na$; they are $(1.03 \pm 0.18) \times 10^{-12}$ and $(1.41 \pm 0.05) \times 10^{-9}$, respectively. The correction due to the geometric differences between the silicon sample and standards isn't applied. Thus, the quantified mass fractions could be affected by a few tens of percents relative error. Moreover, as above reported, the $^{24}Na$ isotope could originate both from $^{23}$Na through $^{23}$Na (n, γ) $^{24}$Na and from $^{28}$Si through $^{28}$Si (n, αp) $^{24}$Na. Therefore, the only quantified contamination of the WASO 04 sample concerned Au.

| Element | Standard | $t_{1/2}$ | Reaction | Peak / keV | Atomic fraction | Mass fraction |
|---|---|---|---|---|---|---|
| As | ML | 1.1 d | $^{75}$As (n, γ) $^{76}$As | 559.1 | ≤ 4.3 × $10^{-12}$ | ≤ 1.2 × $10^{-11}$ |
| Au | ML | 2.7 d | $^{197}$Au (n, γ) $^{198}$Au | 411.67 | (1.47 ± 0.26) × $10^{-13}$ | (1.03 ± 0.18) × $10^{-12}$ |
| Ba | SJ, BR | 11.5 d | $^{130}$Ba (n, γ) $^{131}$Ba | 496.25 | ≤ 2.1 × $10^{-9}$ | ≤ 1.0 × $10^{-8}$ |
| Ca | SJ, BR | 4.5 d | $^{46}$Ca (n, γ) $^{47}$Ca | 1297 | ≤ 9.2 × $10^{-7}$ | ≤ 1.3 × $10^{-6}$ |
| Cd | ML | 2.2 d | $^{114}$Cd (n, γ) $^{115}$Cd | 527.9 | ≤ 1.2 × $10^{-10}$ | ≤ 4.7 × $10^{-10}$ |
| Ce | LL | 32.5 d | $^{140}$Ce (n, γ) $^{141}$Ce | 145.4 | ≤ 1.4 × $10^{-10}$ | ≤ 7.0 × $10^{-10}$ |
| Co | LL | 5.27 y | $^{59}$Co (n, γ) $^{60}$Co | 1332.49 | ≤ 7.1 × $10^{-11}$ | ≤ 1.5 × $10^{-10}$ |
| Cr | LL | 27.7 d | $^{50}$Cr (n, γ) $^{51}$Cr | 320.1 | ≤ 7.0 × $10^{-10}$ | ≤ 1.3 × $10^{-9}$ |
| Cs | LL | 2.06 y | $^{133}$Cs (n, γ) $^{134}$Cs | 795.82 | ≤ 1.3 × $10^{-11}$ | ≤ 6.2 × $10^{-11}$ |
| Eu | BR | 13.5 y | $^{151}$Eu (n, γ) $^{152}$Eu | 1407.74 | ≤ 2.2 × $10^{-11}$ | ≤ 1.2 × $10^{-10}$ |
| Fe | LL | 44.5 d | $^{58}$Fe (n, γ) $^{59}$Fe | 1099.22 | ≤ 5.3 × $10^{-8}$ | ≤ 1.0 × $10^{-7}$ |
| Hf | LL | 42.4 d | $^{180}$Hf (n, γ) $^{181}$Hf | 482.18 | ≤ 1.9 × $10^{-11}$ | ≤ 1.2 × $10^{-10}$ |
| K | SJ, BR | 12.3 h | $^{41}$K (n, γ) $^{42}$K | 1524.75 | ≤ 3.4 × $10^{-9}$ | ≤ 4.7 × $10^{-9}$ |
| La | ML | 1.7 d | $^{139}$La (n, γ) $^{140}$La | 487.02 | ≤ 2.1 × $10^{-12}$ | ≤ 1.0 × $10^{-11}$ |
| Mo | ML | 2.7 d | $^{98}$Mo (n, γ) $^{99}$Mo | 739.5 | ≤ 4.4 × $10^{-10}$ | ≤ 1.5 × $10^{-9}$ |
| Na | SJ, BR | 14.9 h | $^{23}$Na (n, γ) $^{24}$Na $^{(*)}$ | 1368.59 | (1.72 ± 0.06) × $10^{-9}$ | (1.41 ± 0.05) × $10^{-9}$ |
| Nd | SJ$^{(**)}$ | 11.0 d | $^{146}$Nd (n, γ) $^{147}$Nd | 91.11 | ≤ 2.3 × $10^{-10}$ | ≤ 1.2 × $10^{-9}$ |
| Ni | LL | 70.8 d | $^{58}$Ni (n, p) $^{58}$Co | 810.65 | ≤ 1.5 × $10^{-9}$ | ≤ 3.1 × $10^{-9}$ |
| Rb | LL | 18.6 d | $^{85}$Rb (n, γ) $^{86}$Rb | 1076.73 | ≤ 3.5 × $10^{-10}$ | ≤ 1.1 × $10^{-9}$ |
| Sb | LL | 60.2 d | $^{123}$Sb (n, γ) $^{124}$Sb | 1690.84 | ≤ 5.0 × $10^{-11}$ | ≤ 2.2 × $10^{-10}$ |
| Sc | LL | 83.8 d | $^{45}$Sc (n, γ) $^{46}$Sc | 889.23 | ≤ 1.0 × $10^{-11}$ | ≤ 1.6 × $10^{-11}$ |
| Se | LL | 120 d | $^{74}$Se (n, γ) $^{75}$Se | 121.1 | ≤ 9.7 × $10^{-10}$ | ≤ 2.7 × $10^{-9}$ |
| Sm | SJ | 1.9 d | $^{152}$Sm (n, γ) $^{153}$Sm | 103.18 | ≤ 1.7 × $10^{-13}$ | ≤ 9.2 × $10^{-13}$ |
| Ta | LL | 114.4 d | $^{181}$Ta (n, γ) $^{182}$Ta | 1221.26 | ≤ 2.3 × $10^{-11}$ | ≤ 1.5 × $10^{-10}$ |
| Th | LL | 27.0 d | $^{232}$Th (n, γ, $β^-$)$^{233}$Pa | 311.98 | ≤ 8.2 × $10^{-12}$ | ≤ 6.8 × $10^{-11}$ |
| U | ML | 2.3 d | $^{238}$U (n, γ, $β^-$)$^{239}$Np | 228.2 | ≤ 3.5 × $10^{-12}$ | ≤ 2.9 × $10^{-11}$ |
| W | ML | 1.0 d | $^{186}$W (n, γ) $^{187}$W | 685.73 | $\leq 2.9 \times 10^{-12}$ | $\leq 1.9 \times 10^{-11}$ |
| Yb | SJ$^{(**)}$ | 4.18 d | $^{174}$Yb (n, γ) $^{175}$Yb | 396.33 | $\leq 3.3 \times 10^{-12}$ | $\leq 2.0 \times 10^{-11}$ |
| Zn | SJ, BR | 244 d | $^{64}$Zn (n, γ) $^{65}$Zn | 1115.51 | $\leq 4.3 \times 10^{-9}$ | $\leq 1.0 \times 10^{-8}$ |

$^{(*)}$ Possible interference with $^{28}$Si (n, αp) $^{24}$Na, $^{(**)}$ Not certified

**Table 4.** Impurity contents in the WASO 04 sample. The detection limits are evaluated according to Currie's method. The standard uncertainty of the quantified elements include only the counting statistics.

## 6. Conclusions

A review of the literature data concerning the analysis of silicon materials based on neutron activation showed atomic fractions contaminations smaller than $1 \times 10^{-9}$. This confirms the assumption that a large number of elements are almost absent in silicon crystals nowadays produced by the semiconductor industry. However, experimental evidences of the purity level of the silicon crystals used for the determination of the Avogadro constant are still needed.

In this framework, we developed a relative measurement method based on neutron activation. This method was applied to a sample of the WASO 04 crystal and the fractions of twenty-nine elements were determined. Among them, only the mass fraction of Au, about $1 \times 10^{-12}$, was quantified. The mass fractions of sixteen elements were found to be below $1 \times 10^{-9}$. For the remaining twelve elements, the present analytical procedure allowed only to reach mass fraction detection limits ranging between $1 \times 10^{-9}$ and $1 \times 10^{-5}$.

Future work will be aimed at increasing the number of the detectable impurities, at reducing the detection limits, and at certifying the purity of the $^{28}$Si crystal used to determine $N_A$ with the highest accuracy.

## Acknowledgments

This work was jointly funded by the European Metrology Research Programme (EMRP) participating countries within the European Association of National Metrology Institutes (EURAMET) and the European Union.